\documentstyle[aps,epsf]{revtex}
\newcommand\bra[1]{\left\langle\, #1\, \right|}
\newcommand\ev[1]{\left\langle#1\right\rangle}
\def\IP#1#2{\langle\, #1\, |\, #2\, \rangle}
\def\ket#1{\left|\, #1\, \right\rangle}

\begin{document}
\title{\bf Time-of-arrival in quantum mechanics}
\author{
   Norbert Grot\thanks{e-mail: norbert@phyast.pitt.edu }, Carlo 
   Rovelli\thanks{e-mail:  rovelli@pitt.edu},
   Ranjeet S. Tate\thanks{e-mail: rstate@phyast.pitt.edu } \\
   Department of Physics and Astronomy, University of Pittsburgh,\\ 
   Pittsburgh PA 15260, USA}
\date{\today}
\maketitle

\begin{abstract}

\noindent 
We study the problem of computing the probability for the
time-of-arrival of a quantum particle at a given spatial position.  We
consider a solution to this problem based on the spectral
decomposition of the particle's (Heisenberg) state into the
eigenstates of a suitable operator, which we denote as the
``time-of-arrival'' operator.  We discuss the general properties of
this operator.  We construct the operator explicitly in the simple
case of a free nonrelativistic particle, and compare the probabilities
it yields with the ones estimated indirectly in terms of the flux of
the Schr\"odinger current.  We derive a well defined uncertainty
relation between time-of-arrival and energy; this result shows that
the well known arguments against the existence of such a relation can
be circumvented.  Finally, we define a ``time-representation'' of the
quantum mechanics of a free particle, in which the time-of-arrival is
diagonal.  Our results suggest that, contrary to what is commonly
assumed, quantum mechanics exhibits a hidden equivalence between
independent (time) and dependent (position) variables, analogous to
the one revealed by the parametrized formalism in classical mechanics.
\end{abstract}

\section{Introduction}

Consider the following experimental arrangement. A particle moves 
in one dimension, along the $x$ axis.  A detector is placed in the 
position $x=X$.  Let $T$ be the time at which the particle is 
detected, which we denote as the ``time-of-arrival'' of the particle 
at $X$. Can we predict $T$ from the knowledge of the initial state of 
the particle? 

In classical mechanics, the answer is simple.  Let $x(t; x_0, p_0)$ 
be the general solution of the equations of motion corresponding to
initial position and momentum $x_0$ and $p_0$ at $t=0$. We obtain 
the time-of-arrival $T$ as follows. We invert the function $x=x(t; 
x_0, p_0)$ with respect to $t$, obtaining the function $t(x; x_0, 
p_0)$. The time of arrival $T$ at $X$ of a particle with initial data  
$x_0$ and $p_0$ is then 
\begin{equation}  \label{cl:tgen} T = t(X; x_0, p_0).  \end{equation} 
Two remarks are in order. First, if $t(x; x_0, p_0)$ is multivalued, 
we are only interested in its lowest value, since the particle is 
detected the first time it gets to $X$. Second, for certain values of  
$x_0$ and $p_0$, it may happen that $X$ is outside the range of the 
function $x(t; x_0, p_0)$. This indicates that the detector in $X$ 
will never detect a particle with that initial state. The time-of-
arrival is a physical variable that --in a sense-- can take two kinds 
of values: 
either a real number $T\in {\rm I}\! {\rm R}$, or the value: ``$T$ = 
{\it never}".  Notice that in the latter case, the quantity $T$
formally computed from (\ref{cl:tgen}) turns out to be complex.  
Thus, a complex $T$ from (\ref{cl:tgen}) (for given $X; x_0, p_0$) 
means that the particle with initial data $x_0, p_0$ is never
detected at $X$.

In quantum mechanics, the problem is surprisingly harder.  In this 
case the time of arrival can be determined only probabilistically.  
Let $\pi(T)$ be the probability density that the particle 
is detected at time $T$.  Namely, let 
\begin{equation}
	\int_{T_1}^{T_2} \pi(T) dT
\end{equation}
be the probability that the particle is detected between the time 
$T_1$ and the time $T_2$.   How can we compute $\pi(T)$ from 
the quantum state, e.g. from the particle's wave function $\psi(x)$ 
at $t=0$?

To the best of our knowledge, this question has not received a
complete treatment in the standard literature on quantum
mechanics. The problem of computing the time of detection of a
particle is usually treated in very indirect manners.  For instance,
the probability (in time) of detecting a decay product -- say a
particle escaping the potential of a nucleus-- can be obtained from
the time evolution of the probability that the particle is still
within the confining potential.  Alternatively, one can treat the
detector that measures the time-of-arrival quantum-mechanically, and
compute the probabilities for the positions of the detector pointer at
a later time; in this way one can trade a meaurement of the
time-of-arrival for a measurement of position at a late fixed time.
In the fifties, Wigner considered the problem of relating the energy
derivative of the wave function's phase shift to the scattering delay
of a particle \cite{wigner}. This approach, later developed by Smith
\cite{smith} and others (see for instance Gurjoy and Coon \cite{g})
gives the average delay, but fails to provide the full probability
distribution of the time of arrival.  (Smith's paper begins with: ``It
is surprising that the current apparatus of quantum mechanics does not
include a simple representation for so eminently observable a quantity
as the lifetime of metastable entities.'') In the seventies, Piron
discussed the problem in a conference proceeding \cite{piron},
sketching ideas related to the ones developed here.  Ideas related to
the ones presented here were explored in ref. \cite{recami}, but in
this case too only the average time-of-arrival was obtained, and not
its full probability distribution.  Kumar \cite{nkumar} studied the
quantum first passage problem in a path integral approach, but did not
obtain a positive probability density.  The problem has been studied
in the framework of Hartle's generalized quantum mechanics \cite{hartle} 
by using sum over histories methods.  Various attempts in this
direction and discussions of difficulties can be found in Ref.\  
\cite{hallref}. See also the recent paper ref.\ \cite{hartle2} for a 
discussion of the problem and for references; in particular, ref. 
\cite{hartle2} discusses the difficulties one has to face in trying 
to compute {\it sequences} of times-of-arrival -- an important problem 
which, however,  we do not address here.  As stressed by Hartle in 
\cite{hartle2},  generalized quantum mechanics {\it 
generalizes} ``usual  quantum mechanics'';  here, on the other hand, we 
are interested in the question whether $\pi(T)$ can be computed within 
the mathematical framework of  conventional {\it Hamiltonian\,} quantum 
mechanics.

We see two reasons of interest for discussing the problem of computing
time-of-arrival in quantum mechanics.  First, it is a well posed
problem in simple quantum theory, and there must be a solution.
Echoing Smith \cite{smith}, we do not expect that quantum mechanics
could fail to predict a probability distribution that can be
experimentally measured by simply placing a detector at a fixed
position and noting the time at which it ``clicks''.  The problem is
not just academic: it is related to the problem of computing the full
probability distribution (as opposed to the expectation value) for the
tunnelling time through a potential barrier. This problem has
relevance, for instance, in computing rates of chemical reactions
(see, for example, Kumar \cite{nkumar}).  Second, the problem bears
directly on the interpretation of quantum theories without Newtonian
time \cite{carlo_time,isham} and thus on quantum gravity; we shall
briefly comment on this issue in closing.

This paper is the first of a sequence of two.  Here we develop a 
general theory for the time-of-arrival operator, and study the free 
nonrelativistic particle case in detail.   In a companion paper 
\cite{two}, we investigate a technique for the explicit construction 
of the time-of-arrival operator in more general cases, we extent our 
formalism to parametrized systems, and we study some less trivial 
models: a particle in an exponential potential and a 
cosmological model. 

In the next section we give a general argument, based on the 
superposition principle, for the existence of an operator $\hat T$ 
(the time-of-arrival operator) such that $\pi(T)$ can be obtained 
from the spectral decomposition of $\psi(x)$ in eigenstates of $\hat 
T$, in the usual manner in which probability distribution are 
obtained in quantum theory.   $\hat T$ has peculiar properties that 
distinguish it from conventional quantum observables. We give a 
general argument based on the correspondence principle indicating 
that $\hat T$ can be expressed in terms of position and momentum 
operators by the inverse of the classical equations of motion, eq.\  
(\ref{cl:tgen}).  This does not suffice in fixing the operator, since 
factor-ordering ambiguities can be serious.  The problem of the
actual construction of the operator $\hat T$ in more general
systems will be addressed in \cite{two}. In Section 3 we study 
an explicit form of the operator in the case of a free nonrelativistic 
particle. We diagonalize the operator, providing a general expression
for the time-of-arrival probability density $\pi(T)$.  In particular, 
we calculate  $\pi(T)$ explicitly for a Gaussian 
wave packet.  In section 4 we discuss some consequences
of our construction. We notice that the existence
of the operator implies that the quantum mechanics of a 
free particle can be expressed in a ``time-representation'' 
basis. We derive time--energy uncertainty 
relations. We conclude in section 5 with a general comment on the 
equivalence between time and position variables suggested by our 
results.  

In the Appendix, we study whether the probability distribution 
we computed is reasonable, by comparing it with the one 
estimated indirectly using the Schr\"odinger current. We find that 
the two agree within second order in the deBroglie wavelength 
of the particle.  The probability computed from the 
Schr\"odinger current cannot be physically correct to all orders 
because it is not positive definite; whether or not the probability 
distribution computed with $\hat T$ is physically correct to all 
orders is a question that can, perhaps, be decided experimentally.

\section{Time-of-arrival: general theory}

\subsection{The incomplete spectral family $P(T)$}

Consider the quantum analog of the experimental situation sketched 
at the beginning of the paper: a particle is in an initial state $\psi$ 
at $t=0$, and a particle detector is placed at $x=X$.  Let $T$ be the 
time at which the particle is detected.  Let $\pi_\psi(T) dT$ be the 
probability that the particle is detected between times $T$ and 
$T+dT$.  Let $\psi$ and $\phi$ be two quantum states such that both 
$\pi_\psi(T)$ and $\pi_\phi(T)$ have support in the interval $I=(T, 
T+\Delta T)$.  Consider the state formed as the linear combination  
$a\psi+b\phi$ (where $a$ and $b$ are any two complex numbers with 
$|a|^2+|b|^2=1$).  

According to the superposition principle, if a measurable quantity 
has a definite value $\lambda_\psi$ when the system is in the state $\psi$ 
and value $\lambda_\phi$ when the system is in the state $\phi$, 
then a measurement of such a quantity in the state  $a\psi+b\phi$ will 
yield either $\lambda_\psi$, or $\lambda_\phi$ (with respective 
probabilities $|a|^2$ and $|b|^2$) \cite{dirac}. If we assume the 
general 
validity of the superposition principle, we must then expect that the 
probability distribution $\pi_{a\psi+b\phi}(T)$ will have support in 
the interval $(T, T+\Delta T)$ as well.  Therefore, the states $\psi$ 
such that $\pi_\psi(T)$ has support on a given interval $I=(T, 
T+\Delta 
T)$ form a linear subspace of the state space. We can therefore 
define a projection operator $P_I$ as the projector on 
such a subspace.   

The superposition principle could fail for the time-of-arrival. 
However, we would be surprised if it did.   Notice that the question 
can (probably easily) be decided experimentally.   Perhaps an 
experiment testing the validity of the superposition principle in this 
contest could have some interest. Here, we assume that the principle  
holds, and thus the projectors $P_{T, T+\Delta T}$ are well defined.

By their very definition, the projectors satisfy $P_IP_J=P_J$ if the 
interval $J$ is contained in the interval $I$, and $P_IP_J=0$ if the 
two intervals are disjoint.  The operators $P_I$ can therefore be 
written in terms of a family of (spectral) projectors $P(T)$ as 
\begin{equation}
	P_{T, T+\Delta T}=\int_T^{T+\Delta T}P(T')dT'. \end{equation}
The (spectral) family $P(T)$ contains all the information needed to
compute $\pi(T)$. Indeed, from the definition given, and using
again the superposition principle, we have easily 
\begin{equation}\label{pi} \pi_\psi(T) =\bra\psi  P(T) \ket\psi. 
\end{equation}
Thus the probability distribution $\pi(T)$ can be obtained in terms
of the spectral family $P(T)$, in the same way in which all
probability  distributions are obtained in quantum mechanics. Indeed, 
recall that if $\hat A$ is the self-adjoint quantum operator 
corresponding to the observable quantity $A$, then the probability 
distribution $\pi_\psi(A)$ of measuring the value $A$ on the state 
$\psi$ is $\bra\psi P(A) \ket\psi$, where $P(A)$ is the spectral 
family associated to $\hat A$, namely 
\begin{equation} \label{pa}  \hat A=\int A\ P(A)\ dA.  
	\end{equation}

\subsection{The operators $\hat T$ and $\hat {\bf P}$}

One may be immediately tempted to define a ``time-of-arrival
operator'' in analogy with (\ref{pa}) as
\begin{equation}\label{t1} \hat{T} =\int_{-\infty}^{+\infty} T P(T) 
dT \end{equation}
so that an eigenstate of this operator with eigenvalue $T$ would be 
a (generalized) state detected precisely at time $T$. However, there is an
important difference from usual self-adjoint quantum mechanical 
observables that must be addressed before doing so.  If $P(A)$ is the 
spectral family of a self-adjoint operator $\hat A$, then 
\begin{equation}
	\int_{-\infty}^{+\infty} P(A)\ dA = \hat1, \end{equation}
where $\hat 1$ is the identity operator.  On the other hand, 
define $\hat{\bf P}$ by
\begin{equation}\label{projop}
	\hat{\bf P}:=\int_{-\infty}^{+\infty}  P(T)\ dT; \end{equation}
there is no reason for $\hat{\bf P}$ to be the identity. If it is
not, we say that the spectral family $P(T)$ is ``incomplete".

Incompleteness occurs because it is not true that any state is 
certainly detected at some time.  Most likely, there are states that
are never detected, given that such states exist in the
classical theory as well.  Thus, $\hat{\bf P}$ projects on
the subspace ${\cal H}_{\rm detected}$ formed by the states in 
which the particle is detected at some time at $X$, and $(\hat{1}-
\hat{\bf P})$ is the projector on the subspace ${\cal H}_{\rm 
never\,detected}$ of states in which the particle is never detected 
at $X$. The fact that those two classes of states form orthogonal 
linear subspaces follows from the superposition principle again.

Thus, $\hat{T}$ is properly defined by (\ref{t1}) on ${\cal
H}_{\rm detected}$ only. If we define the time-of-arrival operator 
by (\ref{t1}) on the entire state space, then we have the awkward 
consequence that the states in the range of $P(T=0)$ and the ones in 
${\cal H}_{\rm never\, detected}$ are both annihilated by
$\hat{T}$.  Namely $\hat{T}$ does not distinguish the states in
which the particle is detected at $T=0$ from the ones in which it is
never detected.  

The full information that we need in order to compute $\pi(T)$ is 
contained in the incomplete spectral family $P(T)$, or, equivalently, 
in the two mutually commuting operators $\hat{\bf P}$, and 
$\hat{T}$, where $\hat{T}$ is a self-adjoint operator on the Hilbert 
space ${\cal H}_{\rm detected} = Range(\hat{\bf P}(x))$. 

Notice that 
\begin{equation}
\langle \hat{T}\rangle_\psi = \frac{\bra\psi \hat{\bf
P}\hat{T}(x)\hat{\bf P}\ket\psi}{\bra\psi \hat{\bf 
P}\ket\psi}
\end{equation}
is the expected time-of-arrival in those states that are 
detected at all, and is thus a {\it conditional} expectation value.

As defined in (\ref{t1}), $\hat T$ annihilates all states in ${\cal 
H}_{\rm never\ detected}$. It is useful to replace this definition 
by fixing the 
following convention for the action of $\hat T$ on ${\cal H}_{\rm 
never\ detected}$. We define $\hat{T}$ on the {\it entire\,} state 
space, by mimicking what happens in classical mechanics: we choose 
(arbitrarily, at this stage) a (diagonalizable) action of $\hat T$ on 
${\cal H}_{\rm never\ detected}$ with a complex (non-real) 
spectrum, with the understanding that any complex eigenvalue be 
interpreted as ``the particle is never detected".  If we use this 
convention, the operator $\hat T$ is not self-adjoint, but it still 
has a complete and orthogonal basis of eigenstates.  The reason 
for such a convention (which we postulate from now on) will be 
given below; its utility will be particularly clear in \cite{two}.

\subsection{Partial characterization of $\hat T$ from its classical 
limit}

In the previous section we have argued on general grounds that a 
time-of-arrival operator $\hat T$ giving the time-of-arrival 
probability distribution should exist. How can we construct the 
operator $\hat T(x)$, from the knowledge of the dynamics of the 
system?  Let us work in the Heisenberg picture. The quantum theory 
is defined by the Heisenberg state space, in which states do not 
evolve. Let $\psi$ be a Heisenberg state.  The elementary operators 
are Heisenberg position operator $\hat x_0$ and the momentum 
operator $\hat p_0$, representing position and momentum at $t=0$.  
Since all operators can be constructed in terms of
$\hat x_0$ and $\hat p_0$, we expect to be able to express $\hat
T$ as an operator function of $\hat x_0$ and $\hat p_0$.  A key
requirement on $\hat T$ is that it yield the correct results in
the classical limit (Bohr's correspondence principle).  If so, the
dependence of $\hat T$ on $\hat x_0$ and $\hat p_0$ should 
reduce to the classical dependence of $T$ on $x_0$ and $p_0$ in the 
classical limit.  This indicates that the dependence of the operator
$\hat T$ on $\hat x_0$ and $\hat p_0$ is given by some 
ordering of the function (\ref{cl:tgen}), which, we recall, was
obtained by inverting the solutions $x(t; x_0, p_0)$ of the
classical equations of motion.  Thus, we should have that
\begin{equation}
	\label{t11} \hat T = t(X; \hat x_0, \hat p_0) 
\end{equation}
where an ordering has to be chosen. The $c$-number $X$, we recall, 
is the position of the detector.  Notice at this point the usefulness 
of the convention that complex eigenvalues represent non-detection: 
this can go through naturally in the classical limit.  Eq.\ (\ref{t11}) 
does not suffice in general for characterizing $\hat T$ uniquely, 
because the correct physical ordering of the operator 
function can be highly non-trivial. In the companion paper 
\cite{two}, we investigate a technique for constructing the 
operator $\hat T(X)$ and fixing the ordering ambiguities. 

In order to emphasize the dependence of the time-of-arrival 
operator $\hat T$ on the position $X$ of the detector, we will, from 
now on, write the operator as $\hat T(X)$.  Analogously, we will 
denote the spectral family of projectors associated to $\hat T(X)$ 
as $P(T; X)$, and the probability distribution of the 
time-of-arrival at $X$ as $\pi(T; X)$. 

The construction above can be easily generalized to a systems with 
$n$ degrees of freedom.  A classical state of such a system is 
described by a point in the $2n-$dimensional phase space $\Gamma$, 
with coordinates $\{z^i,i=1\cdots 2n\}$.  The dynamics is generated 
by the Hamiltonian $H(z^i)$. The Hamilton equations of motion are $d 
z^i(t)/d t=\{z^i,H\}$.  The general solutions to these equations can be 
written as
\begin{equation} \label{ham}
	z^i(t)=z^i(z^i_0,t),\>   i=1\cdots 2n, \end{equation}
where $z^i_0$ are $2n$ independent integration constants.  In
particular, we may choose as integration constants $z_0^i$ the 
values of $z^i$ at $t=0$, $z^i(0)$.  The above equations enable us to 
compute the state of the system at any time $t$.  We are interested 
in the time-of-arrival of the particle at a given value of {\it one\,} 
of the phase space coordinates, say $z^1$. (Since the coordinates 
$z^i$ are arbitrary, $z^1$ can be any combination of dynamical 
variables.) To compute the time-of-arrival $T(Z)$ at $z^1=Z$, we 
solve the 1st equation of the system (\ref{ham})
\begin{equation}  z^1(z^i_0,t)=Z \end{equation}
with respect to $t$, obtaining $t(Z; z^i_0)$. The time-of-arrival
$T(Z)$ is then given by
\begin{equation}
T(Z)=t(Z; z^i_0). 
\label{t21}
\end{equation}

Now, in the quantum theory, the constants of motion $Z^i_0$ 
correspond to Heisenberg operators $\hat{z}_0^i$.  Eq. (\ref{t11}) is 
immediately generalized by ``quantizing'' (\ref{t21}) as
\begin{equation}\label{t22}
\hat T(Z) = t(Z; \hat{z}_0^i). 
\end{equation}
where, again, the operator $\hat T(Z)$ is given only up to the 
ordering. 

Before concluding this section, we add an important comment on 
the seemingly puzzling case $T<0$, namely when the detection time 
is  {\it earlier\,} than $t=0$.  In the classical case, the particle can 
be detected without being disturbed, but not so in quantum 
mechanics; therefore one might wonder about the meaning of a 
detection at $T<0$ for a particle that has a certain state at time 
$t=0$.  The difficulty is avoided by choosing the definition of 
``state" appropriate to the present context. Consider the classical 
case first.  $x_0$ and $p_0$ fix a unique solution of the equations of 
motion. This solution could be characterized by the values of $x$ and 
$p$ at any other time, or any two constants of the motion. The 
definition of ``time-of-arrival" that avoids the problem of 
detection-before-preparation is the following. We are interested in 
the arrival time of a particle which is moving according to the 
(unique) solution of the equations of motion characterized by the 
fact that at $t=0$ the particle is at $x_0$ and $p_0$ if not 
disturbed.  Analogously, in quantum mechanics $T$ is the 
time-of-arrival of a particle that at an earlier time $t$ 
(arbitrarily in the past) was in the (Schr\"odinger)
state $\psi(t)$ uniquely characterized by the fact that, if not
disturbed, it would evolve to the state $\psi(0)$ at $t=0$.  In the
last section we shall describe a general way of dealing with 
this situation.

\vskip.5cm

To summarize: In this section we have put forward two physical 
hypotheses: 
\begin{itemize}
\item The probability for the time-of-arrival
$\pi(T; X)$ --an experimentally measurable quantity-- can be 
computed by
\begin{equation}
		\pi_\psi(T; X) =\bra\psi  P(T; X) \ket\psi, 
\end{equation}
where $P(T; X)$ are the projectors 
on the real component of the spectrum of a diagonalizable operator 
$\hat T(X)$.  The states in the span of the non-real component
of the spectrum of $\hat T(X)$ are never detected at $X$. 
\item The operator $\hat T(X)$ is given by a suitable 
choice of ordering from the equation  
\begin{equation}
			\hat T(X) = t(X; \hat x_0, \hat p_0) 
\end{equation}
or, in general, Equation (\ref{t22}).
\end{itemize}
The first hypothesis is motivated by our confidence in the general
validity of the superposition principle.  The second hypothesis is
motivated by our confidence in the correspondence principle. In the
next section we investigate some of the implications of these 
hypotheses and we illustrate the construction and the use of the 
operator $\hat T(X)$ in a simple case.  More interesting models, with 
a non trivial $\hat {\bf P}$ operator, will
be presented in the companion paper \cite{two}.

\section{Time-of-arrival of a free particle}

Consider a nonrelativistic free particle in one dimension. Dynamics 
is generated by the Hamiltonian $H:=p^2/2m$.  The solutions of the
classical equations of motion are
\begin{equation}\label{cl}
	x(t; x_0, p_0) = {p_0\over m} \ t + x_0
\end{equation}
The inversion of these yields the time at which a particle that at
$t=0$ has initial position and momentum $x_0, p_0$ is detected
at the position $X$ (as in equation (\ref{cl:tgen}))
\begin{equation}\label{T}
	T(X) = t(X; x_0, p_0) =  {m(X - x_0)\over p_0}.
\end{equation}
Notice that up to problems at the point $p_0=0$ (problems with 
which we shall deal extensively later) the particle is always 
detected.  In particular, $T(X)$ is never complex. This simplifies the 
setting greatly, since we may disregard the complications arising 
from the existence of (finite) regions of phase space in which the 
particle is not detected.

Let us consider the usual quantum theory of a free particle.  We work
in the Heisenberg picture. We have Heisenberg (non-evolving) states
$\psi$, and time dependent Heisenberg position and momentum 
operators $\hat x(t), \hat p(t)$, expressed in terms of $\hat x_0$ 
and $\hat p_0$.   Following the ideas of the previous section, we 
explore the hypothesis that the quantum probability  distribution 
$\pi(T; X)$ of the time-of-arrival at $X$ of the particle can be 
computed in terms of an operator $\hat T(X)$ defined by a suitable 
ordering of the (formal) operator function
\begin{equation} 
	\hat T(X) =  {m(X - \hat x_0)\over \hat p_0}.
\label{eccolo}
\end{equation}

Notice that the Heisenberg position operator is 
\begin{equation}  
\hat x(t) := e^{i\hat Ht/\hbar}\hat x_0 e^{-i\hat Ht/\hbar}
	= {\hat p_0\over m} \ t + \hat x_0;
\label{simple}
\end{equation}
to be compared with (\ref{cl}):  Thought rarely emphasized, 
classical and quantum dynamics are generically related by the 
equation 
\begin{equation}
\hat x(t) = x(t; \hat x_0, \hat p_0)  
\label{dotting}
\end{equation}
where the RHS is an operator function corresponding to an ordering 
of the solution $x(t;  x_0, p_0)$ of the classical equations of motion. 
In general, equation (\ref{dotting}) is of scarce use for solving the 
quantum dynamics, since the associated ordering problem is serious; 
but in simple cases such as the free particle, we see from 
(\ref{simple}) that the natural ordering suffices. 

We explore here the possibility that in the case of
a free particle a natural ordering suffices for the time-of-arrival
operator as well. Namely, we study the choice of a symmetric
ordering for the operator (\ref{eccolo}). We thus 
define, tentatively, 
\begin{equation}\label{t3}
	\hat T(X) :=  {mX\over\hat p_0} - m{1\over\sqrt{\hat p_0}} 
	\hat x_0  {1\over \sqrt{\hat p_0}}. \end{equation}

In order to study this operator, let us choose a concrete
representation for the Hilbert space, namely a basis. It is convenient
to use the momentum basis that diagonalizes $\hat p_0$, because 
this basis makes the definition of ${1\over \sqrt{\hat p_0}}$ 
simpler. Thus, we work in the Heisenberg-picture momentum basis.  
The states are represented as functions $\psi(k)\in L^2({\rm I}\!{\rm 
R})$, and the elementary operators $\hat x_0$ and $\hat p_0$ are 
given by
\begin{equation}\label{momrep}
\hat x_0\circ\psi(k)=i\frac{d}{d
k}\psi(k)\qquad\mbox{and}\qquad \hat p_0\circ\psi(k)=\hbar 
k\  \psi(k). \end{equation}
In terms of the above operators, we have, for example, the 
Heisenberg position operator (\ref{simple})
\begin{equation}
	\hat x(t) = {\hbar k \over m} \ t + i  {d\over d k}.
\end{equation}
In this representation, the operator $\hat T(X)$ given in (\ref{t3}) is
\begin{equation} \label{tonex}
\hat{T}(X)\circ\psi(k) =\left[-i \frac{m} \hbar \frac1 { \sqrt{k}}
\frac{d}{dk}\frac1{\sqrt{k}} +\frac{m}\hbar\frac{X}{k}\right]\psi(k). 
\end{equation}
(We always take the principal value of the square root:
$\sqrt{k}=i\sqrt{|k|}$ for $k<0$.)  

Notice that the $1-$parameter
family of operators $\hat T(X)$ can be generated unitarily via
translations 
\begin{equation}\label{txt0} \hat{T}(X)=e^{-ikX}\hat{T}(0)e^{ikX}.  
\end{equation}
Therefore it is sufficient to study the operator $\hat T(0)$, namely 
we do not loose generality by assuming the detector to be at the 
origin. We thus put $X=0$ from now on, and drop the explicit $X$ 
dependence
\begin{equation} \hat{T} := \hat{T}(0) = -i \frac{m}{\hbar} \frac1 { 
\sqrt{k}}
\frac{d}{dk}\frac1{\sqrt{k}}.  \end{equation} 
We will be interested in the operators corresponding to other
positions of the detector later on.

In the momentum representation, the eigenvalue equation for $\hat 
T$ 
\begin{equation}
\hat T \ket{T}  = T  \ket{T},
\end{equation}
becomes 
\begin{equation}
\hat {T}\circ g_T(k)\equiv 
\left[-i\frac{m}\hbar\frac1{\sqrt{k}}\frac{d}{dk}\frac1{\sqrt{k}}
\right]g_T(k)= T\   g_T(k).
\end{equation}
where we have introduced the notation 
\begin{equation}
g_{T}(k)\equiv\IP{k}{T} 	
\end{equation}
for the momentum representation of the eigenstate of $\hat T$.   
The eigenvalue equation is easily solved (in each half of the real line 
($k\ne0)$) by
\begin{equation}
g_T(k)=\left(\alpha_+\,\theta(k)+\alpha_-\,\theta(-k)\right)
\sqrt{\frac\hbar{2\pi m}}\sqrt{k}\exp\left(\frac{i\hbar T
k^2}{2m}\right),
\end{equation}
where $\theta(k)$ is the characteristic function of the positive half
of the real line, and $\alpha_\pm$ are
constants independent of $k$.
In order to fix the relation between $\alpha_+$ and $\alpha_-$, let 
us act on $g_T(k)$ by $\hat T$. A simple calculation shows
\begin{equation}\hat{T}\circ g_T(k)=T\cdot g_T(k)-i 
\sqrt{\frac{m}{2\pi\hbar
}}\,\frac{\delta(k)}{\sqrt{|k|}}\> (\alpha_+ +i\alpha_-).
\end{equation}
Thus, in order to satisfy the eigenvalue equation, it is necessary 
that%
\footnote{Another approach to obtaining this result is to integrate
the eigenvalue equation in a small region around $k=0$. One then
obtains the same continuity condition (\ref{concon}) on $g_T(k)$.}
\begin{equation}\label{concon}
		\alpha_-=i\alpha_+.
\end{equation}
At this point, we encounter a difficulty. The operator we have
constructed does {\em not} have a basis of orthogonal eigenstates.  
This pathology destroys the possibility of giving $\hat T$ the 
interpretation we want.  In the next subsection 
we show that the eigenstates of $\hat T$ are not orthogonal and we 
discuss a way out from this difficulty.

\subsection{Difficulties with $\hat T$ and a regulation}

A simple calculation shows that for any two
eigenstates of $\hat T$ with eigenvalues $T$ and $T'$ 
\begin{equation}\IP{T}{T'}\equiv\int_{-\infty}^\infty 
dk\>\overline{g_T(k)}g_{T'}(k)
=\frac\hbar{2\pi m}\int_0^\infty
dk^2\>e^{\frac{i\hbar}{2m}k^2(T'-T)}=\delta(T-T')-\frac{i}{\pi(T-T')}.
\end{equation} 
The eigenstates fail to be orthogonal. One can also see that $\hat
T$ as defined above has no self-adjoint extensions by noticing that
its deficiency indices are unequal.

This difficulty stalled us for sometime, and various attempts 
to
circumvent the problem failed.  A way out was then suggested by 
Marolf
\cite{dm:pc}.  The idea is to seek an operator that in the classical
limit would not reproduce the time-of-arrival exactly, but would 
rather
reproduce a quantity arbitrary close to the time-of-arrival.  Namely,
we want to approximate the time-of-arrival with a different 
quantity,
free from pathologies.  It is easy to trace the above pathology to the
singular behavior of $1/k$ at $k=0$. Even classically, a state with
$k=0$ is physically disturbing: either the particle is never detected
or the particle may stably sit over the detector.  Therefore, we seek
a small modification of (\ref{T}) such that no divergences occur at
$p_0=0$.  The modified time-of-arrival can perhaps be interpreted 
as the outcome of a measurement by an apparatus 
arbitrarily similar to a perfect detector, but which does not allow 
the particle to stand still.

Let us introduce an arbitrary small positive number $\epsilon$. 
Consider a 1--parameter family of real bounded continuous odd 
functions ${f_\epsilon}(k)$ which approach $1/k$ pointwise. 
More precisely, we require   
\begin{eqnarray}
{f_\epsilon}(k) &=& {1 \over k} \qquad {\rm for} \qquad |k| > 
\epsilon\cr
&&\mbox{and nonzero for all}\quad k\ne0. 
\end{eqnarray}
For instance, we may choose 
\begin{eqnarray}\label{f2} 
{f_\epsilon}(k) &=& {1 \over k} \qquad\ \ \  {\rm for} \qquad |k| > 
\epsilon, \cr {f_\epsilon}(k) &=& \epsilon^{-2} k \qquad {\rm for} 
\qquad
|k| < \epsilon.  
\end{eqnarray}
Using this, we define the regulated time-of-arrival operator as 
\begin{equation}\label{regt1} \hat T_\epsilon =
 -i\frac{m}{\hbar}\sqrt{f_\epsilon(k)}\frac{d}{d k}
 \sqrt{f_\epsilon(k)}, 
\end{equation}
to be compared with the unregulated operator 
(\ref{tonex}).  Notice that on any state with 
support on $|k|>\epsilon$ the operators $\hat T_\epsilon $ 
and $ \hat T$ are equal.  
Their action differs only on the component of a state 
with arbitrary low momentum. 
As we shall see, the probability distribution 
for the time-of-arrival  $\pi(T)$ computed from $\hat T_\epsilon$  
will turn out to be {\it 
independent of $\epsilon$} for states with support away from
$k=0$  --reinforcing the credibility of the 
regulation procedure we are using.  

Let us study the operator $\hat T_\epsilon$.  A key point is that 
$\hat T_\epsilon$ commutes with $\Theta(k)=sgn(k)=\frac{k}{|k|}$. 
Thus, we can choose a basis of solutions of the eigenvalue equation
for $\hat T_\epsilon$ 
formed by functions of $k$ which have support 
on positive or negative $k$ only. Now $\hat 
T_\epsilon$ is a linear differential operator and
since $f_\epsilon(k)\rightarrow0$ as $k\rightarrow0$, there is no 
continuity condition on its eigenstates at $k=0$.  These two related
properties lead to a degeneracy in the spectrum. For each
eigenvalue $T$, there are two eigenstates, which we choose as 
having support in the $k>0, k<0$ regions respectively. Namely
\begin{eqnarray}
\hat T_\epsilon \ket{T, +}_\epsilon  &=& T  \ket{T, +}_\epsilon ,
\nonumber \\  
\hat T_\epsilon \ket{T, -}_\epsilon  &=& T  \ket{T, -}_\epsilon , 
\end{eqnarray}
where
\begin{eqnarray}
\IP{k}{T, +}_\epsilon = &0 & \ \ \ {\rm for}\ k<0,
\nonumber \\  
\IP{k}{T, -}_\epsilon = &0 & \ \ \ {\rm for}\ k>0. 
\end{eqnarray}
We introduce the notation 
\begin{equation}
{{}_\epsilon g}^\pm_{T}(k)\equiv\IP{k}{T,\pm}_\epsilon 	
\end{equation}
for the momentum representation of the eigenstates of $\hat 
T_\epsilon$.   A simple calculation shows that these are
\begin{equation}
\label{gt0}
{{}_\epsilon g}^\pm_{T}(k) = \theta(\pm k) \sqrt{\frac\hbar{2\pi
m}}\frac1{\sqrt{{f_\epsilon}}} \exp\left(i\frac{\hbar
T}{m}\int_{\pm\epsilon}^k dk' ({f_\epsilon}(k'))^{-1}\right). 
\end{equation}
Explicitly, for $|k|\ge \epsilon$ we have
\begin{equation} 
\label{stati}
{{}_\epsilon g}^\pm_{T}(k)= \theta(\pm k) \sqrt{\frac\hbar{2\pi 
m}}\sqrt{k}
\exp\left(\frac{i\hbar T}{2m}(k^2-\epsilon^2)\right).  
\end{equation}

In order to derive various properties of these eigenstates, it is 
convenient to introduce new coordinates on the right and left halves 
of the real line, as follows.
\begin{equation} 
	Z^\pm(k)=\int_{\pm\epsilon}^k dk'\frac1{{f_\epsilon}(k')}.
\end{equation}
In the region $|k|>\epsilon$, $Z^\pm(k)=(k^2-\epsilon^2)/2$.  In what
follows, we do not need the specific form of $Z^\pm$ in the region
$|k|<0$. Note first that in each
half, the Jacobian of the coordinate transformation is non-vanishing
$\forall k\ne0$, and thus, the new coordinates are strictly
monotonic.  At the points $|k|=\epsilon$, $Z^\pm(k=\pm\epsilon)=0$
respectively. Furthermore, since $|{f_\epsilon}(k)|\rightarrow0$ 
rapidly
enough as $|k|\rightarrow0$, we see that both
$Z^\pm\in(-\infty,\infty)$. In terms of these coordinates, the
eigenstates are
\begin{equation}\label{epeigz}
{{}_\epsilon g}_T^\pm(Z^\pm)=\sqrt{\frac\hbar{2\pi 
m}}\frac1{\sqrt{{f_\epsilon}
(k(Z^\pm))}}e^{i\frac\hbar{m}T\cdot Z^\pm}
\end{equation}
and the Hermitian inner product between two
states $\psi,\phi$ is
\begin{equation}\label{zip}\IP\psi\phi=\sum_{\eta=+,-}\int_{-
\infty}^\infty 
dZ^\eta{f_\epsilon}(k(Z^\eta))\overline{\psi(k(Z^\eta))}\phi(k(Z^\eta)
) ,
\end{equation}
From
(\ref{epeigz},\ref{zip}), the 
orthogonality of the eigenstates is manifest:
\begin{equation}
{}_\epsilon\!\IP{T,\eta}{T',\eta'}_\epsilon
=\delta_{\eta,\eta'}\delta(T-T').
\end{equation}
In the new coordinates, completeness too is manifest. We can get 
the
same result in the $k$-representation with a little work:
\begin{eqnarray}\label{complete}
\sum_{\eta=+,-}\int_{-\infty}^\infty
dT\IP{k}{T,\eta}_\epsilon\>{}_\epsilon\!\IP{T,\eta}{k'}&=&
\frac\hbar{2 \pi m}\sum_\eta\int_{-\infty}^\infty dT
\frac1{\sqrt{{f_\epsilon}(k)}}
\frac1{\overline{\sqrt{{f_\epsilon}(k')}}}e^{i\frac\hbar{m}T(Z^\eta(k
)-
Z^\eta(k'))}\cr
&=&\sum_\eta\frac{\delta(Z^\eta(k)-Z^\eta(k'))}
{\sqrt{{f_\epsilon}(k)}\overline{\sqrt{{f_\epsilon}(k')}}}=
\sum_\eta\frac{\delta(k-k')\theta(\eta k)}
{|\partial Z^\eta/\partial 
k|\>\sqrt{{f_\epsilon}(k)}\overline{\sqrt{{f_\epsilon}(k')}}}\cr
&=&\delta(k-k').
\end{eqnarray}
Since it has 
a complete orthogonal basis of (generalized) eigenstates 
with real eigenvalues, $\hat T_\epsilon$ is self-adjoint.

\subsection{Time-of-arrival probability density}

Following the general theory of section 2, if the particle is in the 
Heisenberg state $\psi(k)$, the probability density $\pi(T)$ 
of the time-of-arrival is the modulus square of the projection of
the state on the $T$-eigenstates of the time-of-arrival operator. 
Since these are doubly degenerate, we have in the present case
\begin{equation}\label{probdensT}
	\pi(T)= |{}_\epsilon\IP{T,+}\psi|^2+|{}_\epsilon\IP{T,-}\psi|^2. 
\end{equation}
If we assume that the support of $\psi(k)$ does not contain (an 
arbitrary small finite region $|k|<\delta$ around) the origin, we can 
choose $\epsilon<\delta$ and, using the explicit form (\ref{stati}) 
of the eigenstates, we obtain the following expression for 
$\pi(T)$
\begin{equation}\label{probdensT2}
\pi(T)=\frac\hbar{2 \pi m}  \left(
\left|\int_0^{\infty}
  dk \sqrt{k}\exp\left(\frac{iT\hbar (k^2-\epsilon^2)}{2
m}\right)\psi(k) \right|^2
+  
\left|\int^0_{-\infty}
  dk \sqrt{k}\exp\left(\frac{iT\hbar (k^2-\epsilon^2)}{2
m}\right)\psi(k) \right|^2\right).
\end{equation}
Notice that the $\epsilon$ dependence gives only a phase that 
disappears when we take the absolute value squared. 
Namely
\begin{equation}\label{probdensT3}
\pi(T)=\frac\hbar{2 \pi m} 
\left(\left|\int_0^{\infty}
  dk \sqrt{k}\exp\left(\frac{iT\hbar k^2}{2
m}\right)\psi(k) \right|^2
+  
\left|\int^0_{-\infty}
  dk \sqrt{k}\exp\left(\frac{iT\hbar k^2}{2
m}\right)\psi(k) \right|^2\right).
\end{equation}
We thus have the result that for the states that do not include an 
amplitude for zero velocity, the time-of-arrival probability 
distribution computed  (with $\epsilon$ sufficiently small) with the 
regulated  operator $\hat T_\epsilon(x)$ is independent from 
$\epsilon$. 

The two terms in (\ref{probdensT3}) correspond to the 
left and right moving component of the state. Therefore, we have 
immediately that the probability $\pi^+(T)$ (and $\pi^-(T)$) that the 
particle is detected in $X=0$ while moving in the positive (or  
negative) direction is 
\begin{equation}
\pi^\pm(T)=\frac\hbar{2 \pi m} 
\left|\int_0^{\pm\infty}
  dk \sqrt{k}\exp\left(\frac{iT\hbar k^2}{2
m}\right)\psi(k) \right|^2.
\end{equation}

Finally, the result generalizes immediately to the case in which the 
detector is not placed in the origin, but rather in an arbitrary 
position $X$. The eigenstates of the operator $\hat T_\epsilon(X)$
\begin{equation}
	\hat T_\epsilon(X)  \ket{T, \pm; X}_\epsilon 
= T \ket{T, \pm; X}_\epsilon 
\end{equation}
are obtained using the unitary translation operator
\begin{equation}
	 \ket{T, \pm; X}_\epsilon = e^{-{i\over\hbar}\hat pX} \ket{T, 
\pm}_\epsilon.
\end{equation}
yielding 
\begin{equation} 
{{}_\epsilon g}^\pm_{T
;X}(k) = \IP{k}{T,\pm; X}_\epsilon = e^{-ikX} 
{{}_\epsilon g}^\pm_{T}(k).
\end{equation}
The projectors considered in section 2 are given by 
\begin{equation}
 P(T; X) = \ket{T, +; X}{}_\epsilon\,{}_\epsilon\!\bra{T, +; X}
+\ket{T,-; X}{}_\epsilon\,{}_\epsilon\!\bra{T, -; X}.  
\end{equation} 
The probability density of being detected at time $T$ by a detector 
in $X$ that detects particles traveling with positive (negative) 
velocity is
\begin{equation}
\pi^\pm(T; X)=\frac\hbar{2 \pi m} 
\left|\int_0^{\pm\infty}
  dk \sqrt{k}\exp\left(\frac{iT\hbar k^2}{2
m}-ikX\right)\psi(k) \right|^2.
\label{final}
\end{equation}
Equation (\ref{final}) represents our final result for the probability 
distribution of the time-of-arrival at $X$ of a free quantum 
particle.

\subsection{Time-of-arrival of a Gaussian wave packet}

As an example of an application of the above result, we compute the
probability distribution for the time-of-arrival of a Gaussian wave
packet. Consider a Gaussian wavepacket localized about a point (say)
to the left of the origin at time $t=0$, and moving (say) to the
right.  In the standard Schr\"odinger-picture position representation,
let this wave packet be given by the following normalized solution of
the Schr\"odinger equation
\begin{equation}\label{ss:xt}
\psi(x, t)=\left(\frac{\delta^2}{2 \pi}\right)^{\frac14} 
\frac{e^{-k_0^2\delta^2}}{\sqrt{\delta^2+ \frac{it\hbar}{2m}}} \exp\left(
\frac{(2\delta^2k_0+i(x-x_0))^2}{4\delta^2+2i\hbar t/m} \right).
\end{equation}
Expectation values are as follows
\begin{eqnarray}
\ev{p(t)}=\hbar k_0 &\quad\mbox{ and }\quad& \ev{\Delta
p(t)}=\frac\hbar{2 \delta}\\
\ev{x(t)}=x_0+\hbar k_0t/m &\quad\mbox{ and }\quad& \ev{\Delta 
x(t)}=\delta\sqrt{1 + \frac{t^2\hbar^2}{4\delta^4 m^2}}. 
\end{eqnarray}
If we choose $|x_0|>>\delta$, $|k_0|\delta>>1$, $x_0<0$ and $k_0>0$, 
this state represents a particle well localized to the
left of the origin and with a well-defined positive momentum at 
time $t=0$.  In the Heisenberg-picture momentum representation 
(\ref{momrep}),  this state is given by
\begin{equation}\label{ss:krep}
\psi(k)=\left(\frac{2\delta^2}{\pi}\right)^{\frac14} \exp\left(
-(k-k_0)^2\delta^2 -ikx_0\right).
\end{equation}
The envelope of this wave function is a Gaussian of width 
$1/\delta$ centered at $k_0$.

Using the theory developed, we can compute the projection of this 
state on the eigenstates of the time-of-arrival operator.  We assume 
here that $\epsilon$ can be taken arbitrarily close to $0$. (See
\cite{GandR} for the relevant integrals). 
We obtain 
%
\begin{eqnarray}\label{psi(tx)}
{}_\epsilon\IP{T,+; X}\psi &=&
\sqrt{\frac\hbar{m}}\left(\frac{\delta^2}{2^5\pi^3}\right)^{1/4}
e^{-\delta^2 k_0^2}\>\Gamma(5/4)\times\cr
&&\left(\delta^2+\frac{i\hbar T}{2m}\right)^{-5/4}\left\{
\left((2\delta^2k_0+i(X-x_0))\cdot 
\Phi\left[\frac54,\frac32,\frac{(2\delta^2k_0+i(X-x_0))^2}
{4\delta^2+2i\hbar T/m}\right]\right.\right.\cr
+\sqrt{\pi\left(\delta^2+\frac{i\hbar T}{2m}\right)}
&\times&\left.\left.\exp\left(
\frac{(2\delta^2k_0+i(X-x_0))^2}{4\delta^2+2i\hbar T/m}\right)
\cdot {\cal L}\left[\frac14,-\frac12,-\frac{(2\delta^2k_0+i(X-
x_0))^2}
{4\delta^2+2i\hbar T/m}\right]\right)\right\},\\ 
&&\cr
{}_\epsilon\IP{T,-; X}\psi &=&
\sqrt{\frac\hbar{m}}\left(\frac{\delta^2}{2^5\pi^3}\right)^{1/4}
e^{-\delta^2 k_0^2}\>\Gamma(5/4)(-i)\times\cr
&&\left(\delta^2+\frac{i\hbar T}{2m}\right)^{-5/4}\left\{
\left(-(2\delta^2k_0+i(X-x_0))\cdot 
\Phi\left[\frac54,\frac32,\frac{(2\delta^2k_0+i(X-x_0))^2}
{4\delta^2+2i\hbar T/m}\right]\right.\right.\cr
+\sqrt{\pi\left(\delta^2+\frac{i\hbar T}{2m}\right)}
&\times&\left.\left.\exp\left(
\frac{(2\delta^2k_0+i(X-x_0))^2}{4\delta^2+2i\hbar T/m}\right)
\cdot {\cal L}\left[\frac14,-\frac12,-\frac{(2\delta^2k_0+i(X-
x_0))^2}
{4\delta^2+2i\hbar T/m}\right]\right)\right\},\cr
&&
\end{eqnarray}
where we have ignored the components in $|k|<\epsilon$, 
since $\epsilon$ can
be taken arbitrarily small. ($\Phi[n,m;z]={}_1F_1[n;m;z]$ is the 
Kummer
confluent hypergeometric function, ${\cal L}[n,a;z]={\cal L}_n^a(z)$
is the $n$th generalized Laguerre polynomial and $\Gamma(n)$ is the
Euler gamma function.)   The probability distribution is then given by
(\ref{probdensT}).  
The expression above for the probability distribution of the time of 
arrival of a Gaussian wave packet is a bit heavy; in order to unravel 
its content, we have expanded it in powers of small quantities in the 
Appendix A, and we have plotted 
the total probability 
density $\pi(T)$ at various detector positions $X$ in Figure 1  
(choosing a Gaussian state (\ref{ss:xt}) with
$x_0=-5, k_0=20, \delta=.5, \hbar=m=1$).   To begin with, 
the term corresponding to negative
velocities is exponentially small; 
indeed, it derives from the scalar product of a Gaussian wave packet 
concentrated around a positive $k$ with a function having support on 
$k<0$. 
\begin{figure}
\epsfxsize=400pt\epsfbox{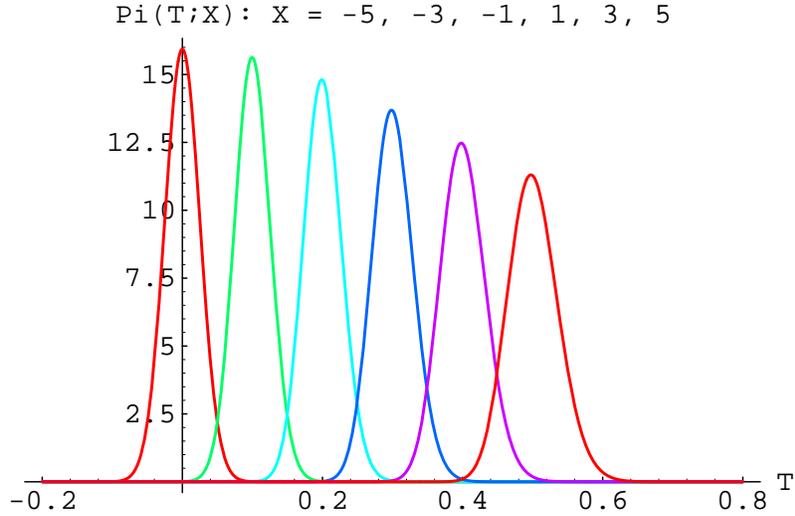}
\caption[Fig. 1.]{Time-of-arrival probability densities $\pi(T; X)$,
plotted at the detector positions $X=-5,-3,-1,1,3,5\>$. }
\end{figure}
The total detection probability density at time $T$ for the detector
in position $X$ is a function (more precisely, it is a density in $T$)
on the $T-X$ plane. This function is concentrated around the classical
trajectory of the particle $X=x_0+ p_0 T/m$, with a (quantum) spread
in $T$ that increases with the spread of the wave packet, namely with
the distance of the detector from the initial state.

In Appendix A, we compare our result with the probability density 
obtained indirectly using the Shr\"odinger probability density. We 
find good agreement within second order in the deBroglie wavelength 
of the  particle, and we discuss the order of the discrepancies.  
Thus, our result is reasonable to leading order. Whether or not it is 
physically correct to all orders is a question that can perhaps be 
decided experimentally.  A discrepancy with an experimental result 
may indicate an incorrect ordering of the time-of-arrival operator, 
or a more general difficulty with our approach.  

\section{Discussion}

\subsection{Time representation}

Anytime we have a self-adjoint operator in quantum mechanics, we 
may define a representation that diagonalises this operator. 
Namely, we may choose the eigenbasis of the operator as 
our working basis on the theory's Hilbert space.  Nothing prevents us 
from doing so with the operators $\hat T_\epsilon(X)$ as well.  Let 
us therefore introduce a ``time-of-arrival representation'', or, for 
short, a ``time-representation''.  We fix an $ \epsilon $ and define 
\begin{equation}\label{tofxrep}
\psi^\pm_\epsilon(T, X):= {}_\epsilon \IP{T,\pm; X}{\psi} =
\int dk \ \overline{{{}_\epsilon g}_{T;X}^\pm(k)} \ \psi(k).  
\end{equation}
Clearly, we can do quantum mechanics in the 
$\psi^\pm_\epsilon(T, X)$
representation, as well as we do quantum mechanics in the position,
momentum, or energy representations.  Since the eigenstates
${{}_\epsilon g}_{T, X}^+(k)$ have support on positive $k$, we must 
interpret
$\psi^+_\epsilon(T, X)$ as the amplitude for the particle to be 
detected by a detector placed at $X$ in an infinitesimal 
neighborhood
of $T$ coming from the left, and $\psi^-_\epsilon(T, X)$ as the
amplitude to be detected at $X$ in a neighborhood of $T$ coming
from the right.

What is the relation between the $\psi^\pm_\epsilon(T, X)$ amplitude
and the conventional Schr\"odinger wave function $\psi(x,t)$?  Notice
that the first is defined by $\psi^\pm_\epsilon(T, X) =
{}_\epsilon\IP{T,\pm; X}{\psi}$, where $\ket{T,\pm; X}{}_\epsilon$ is
an eigenstate of $T_\epsilon(X)$ with eigenvalue $T$; while the second
can be viewed as defined by $\psi(x,t)=\IP{x;t}{\psi}$, where $\ket{x;
t}$ is the eigenstate of the operator $\hat x(t)$ with eigenvalue $x$.
At first sight, the two seem to be related to the same quantity (up to
the $\epsilon$ and the distinction between the two directions of the
velocity): they both refer to probabilities of being detected at some
space-point and at some time.  However, this naive observation is very
misleading.  The quantity $|\psi(x, t)|^2dx$ is the probability {\it
in space} that the particle happens to be between the positions $x$
and $x+dx$ at time $t$, {\it as opposed to being elsewhere\,} at time
$t$.  Whereas, the quantity $|\psi^\pm(T, X)|^2dT$ is the probability
{\it in time} that the particle happens to arrive between times $T$
and $T+dT$ at the position $X$, {\it as opposed to reaching $X$ at
some other time\,}.  The two bases $\ket{T, \pm; X}{}_\epsilon$ and
$\ket{x; t}$ are two well defined (generalized one-parameter families
of) bases in the Hilbert space, but they are distinct.

In particular, the two bases $\ket{T, \pm; X}{}_\epsilon$ and $\ket{x; t}$  
have distinct 
dimensions, because  $|\psi(x, t)|^2\ dx$ and  $|\psi^\pm(T, X)|^2\ dT$ 
must both be dimensionless probabilities.  Thus, the transformation 
factor between $\ket{T, \pm; X}$ and $\ket{x; t}$ has the dimension 
of the square root of a velocity.  Indeed, let us write the two 
(generalized) 
states explicitly in the Heisenberg momentum representation. 
Restricting ourselves to $k>0$ and taking $\epsilon$ to zero for 
simplicity, we 
have from (\ref{stati})
\begin{equation} 
\IP{k}{T, +; X}{}_\epsilon =
\sqrt{\frac\hbar{2\pi}}\sqrt{{k\over m}}
\exp\left(\frac{i\hbar T}{2m}k^2-ikX\right), 
\end{equation}
while, as it is well known, 
\begin{equation} 
\IP{k}{x, t} =
\sqrt{\frac\hbar{2\pi}} 
\exp\left(\frac{i\hbar t}{2m}k^2-ikx\right).
\end{equation}
Therefore, taking $x=X$ and $t=T$ we have
\begin{equation} 
\IP{k}{T, +; X}{}_\epsilon = \sqrt{{k\over m}}\  \IP{k}{x, t}. 
\end{equation}

A physical understanding of the curious $\sqrt{{k\over m}}$ factor 
that characterizes the eigenstates of the time-of-arrival operator 
can be obtained as follows. Consider a well localized wave packet 
travelling with velocity $v=k_0/m$. We have approximately  
$|\psi^\pm(T, X)|^2 \approx v\  |\psi(x, t)|^2$.   Now, consider the $x-t$ 
plane. The wave function $\psi(x, t)$ is significantly different from 
zero on a band around the classical trajectory. The classical 
trajectory is a straight line with a slope given by the velocity 
$v$. The ratio between a vertical and an horizontal section of the 
band is therefore precisely $v$. Thus, in order to have both total 
probabilities normalized to 1 when integrating along a $t$ = 
constant, or a $x$ = constant line, the probability density in space 
and the probability density in time must be related by a factor $v$. 

In Figure 2 we have plotted the usual Schr\"odinger probability
density in $x$ -- $\pi(x;t)=|\psi(x, t)|^2$ -- at various 
times, for the same state used for Figure 1. 
\begin{figure}
\epsfxsize=400pt\epsfbox{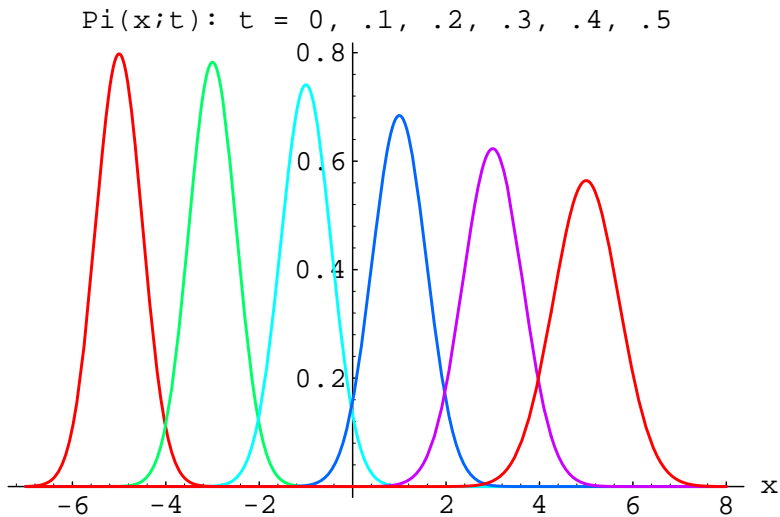}
\caption[Fig. 2.]{Schr\"odinger probability densities $\pi(x; t)$ for 
position,
plotted at the times $t=0,.1,.2,.3,.4,.5\>$. }
\end{figure}
The time-of-arrival probability densities are $v=k_0/m$ 
times as large as the Schr\"odinger densities for values of $X=x$ and 
$T=t$ near the classical trajectory.

\subsection{Time-energy uncertainty relation}

Position-momentum uncertainty relations are of wide use in 
quantum
mechanics, and can be cleanly derived from the formalism on very
general grounds.  The commutation relations $[\hat x, \hat p]=-
i\hbar
\hat 1$ imply $\Delta x \Delta p >2h$.  Time-energy uncertainty
relations are of wide use as well (e.g. between the width of a
spectral line and the lifetime), but their general derivation is
notoriously tricky.  If one tries to reproduce the
position-momentum derivation for the time-energy case by 
assuming the existence of an operator $\hat T$ such that
\begin{equation}
[\hat T, \hat H]=-i\hbar \hat 1 \label{cr} \end{equation}
from which 
\begin{equation}
	\Delta T \Delta E >2h \label{ur} \end{equation}
would follow, then one would clash against a well known 
non-existence theorem for
$\hat T$.   The theorem states that the commutation
relations $[\hat A, \hat B]=-i\hbar\hat1$ between two self-adjoint
operators $\hat A$ and $\hat B$ implies that the spectrum of both
operators is the real line. However, the spectrum of the Hamiltonian
is bounded from below in all reasonable systems. {\it Ergo\,} a time
operator $\hat T$ satisfying (\ref{cr}) does not exist.

This theorem might have been the reason for which a 
time-of-arrival operator has virtually never been considered 
in quantum mechanics. In fact, it is often stated that 
time cannot be an operator in quantum mechanics,  
with the above theorem as a proof. Here, we show how one 
can rigorously derive time-energy uncertainty relations for 
a quantum particle, and how the existence of 
$\hat T_\epsilon$ circumvents the theorem. 

The commutation relations between $\hat T_\epsilon$ and the 
Hamiltonian are easy to compute. In the momentum representation 
we have
\begin{equation}
[\hat T_\epsilon,\hat H] = -i\hbar \hat 1 - i\hbar  h_\epsilon(k), 
\end{equation}
where
\begin{equation}
h_\epsilon(k) = 1 - k f_\epsilon(k). 
\end{equation}
The function $h_\epsilon(k)$ is bounded (by $\epsilon^2$, if we 
chose $f_\epsilon(k)$ as in (\ref{f2}), which we do here for 
simplicity) and has support on the small interval $|k|<\epsilon$.  For 
the particle in the state $\psi(k)$ the following uncertainty 
relations follow
\begin{equation}
	\Delta T_\epsilon \Delta E > 2 h - 4\pi \epsilon^2 
 	\int_{-\epsilon}^{+\epsilon} dk\  |\psi(k)|^2 >  2 h - 4\pi 
\epsilon^2. 
\end{equation}
By chosing $\epsilon$ sufficiently small, we obtain an uncertainty 
relation that approaches (\ref{ur}) to any desired precision. 

\subsection{On the definition of state in quantum mechanics}

Finally, let us return to the problem we briefly discussed at the end 
of section 2, which is the interpretation of the time-of-arrival
$T$ when $T<0$, namely when detection time is earlier than the 
time $t=0$ at which the initial state is given.  We have suggested 
that in this case the correct interpretation of $T$ is the
following. $T$ is the detection 
time for a state that arbitrarily in the past was in a state that 
would have
evolved to the $t=0$ initial state if undisturbed.  A cleaner way of
dealing with the general situation, is to make use of a fully
time-independent notion of ``state'' and a fully time-independent
version of phase space and quantum state space.  This can be done as
follows.  Consider first classical mechanics.  Let us denote a single
solution of the equation of motion as a ``physical history" of the
system.  (A physical history should not be confused with the 
histories considered in sum-over-histories theories: a ``physical
history'' here is a history satisfying the equations of motion.)  Let
$\Gamma_s$ be the space of these physical histories.  A point in
$\Gamma_s$ represents an entire evolution of the system.  
$\Gamma_s$ can be coordinatized by the 2$n$ integration constants 
$z^i$. We ask for the time-of-arrival at $x$ of a system following 
one of the
motions in $\Gamma_s$. This time-of-arrival is given by
(\ref{t21}). The key to the matter is that there is no need to choose
a time in order to specify a physical history.

In quantum mechanics we can define the Hilbert space ${\cal H}_S$ 
of
the {\it solutions,} of the Schr\"odinger equation.  A vector in $H_S$
represents an entire (quantum) motion of the system, without 
reference
to any particular time. The conventional Heisenberg operators are
defined on ${\cal H}_S$.  The operator (\ref{t22}) is properly defined
on ${\cal H}_S$.  We may choose to represent the vectors in ${\cal
H}_S$ by means of the value that the Schr\"odinger state would take
(if undisturbed) at $t=0$; therefore it makes sense to define $\hat 
T$ as a function of the operators $x_0$ and $p_0$ which are defined 
on the states at $t=0$.

In this regard, it is interesting to notice that 
the original definition of the ``Heisenberg picture Hilbert space"
given by Dirac in the first edition of ``{\it Principles of Quantum 
Mechanics}'' is the definition of ${\cal H}_S$ given above 
\cite{dirac}. 
It is only later that the ``Heisenberg picture Hilbert space'' came to 
be mostly identified with the state space at a fixed time (both  
interpretations of ${\cal H}_S$ can be found in the literature).  A 
crucial advantage of using the definition of ${\cal H}_S$ given here 
is that this definition can be extended to systems without 
Newtonian time  at all \cite{carlo_time,deparam}. We will exploit 
this point of view in \cite{two}. 

\section{Conclusions: $x\leftrightarrow t$ equivalence in quantum 
theory}

Let us summarize our results.  We have considered the problem of
computing the time-of-arrival $T$ of a quantum particle at a 
position $X$.  Relying on the general validity of the superposition 
principle, we have argued that the probability distribution for $T$ 
can be obtained by means of an operator $\hat T$.  This operator is, 
in general, not self-adjoint.  However, it admits an orthogonal basis
of eigenstates. The eigenstates with real eigenvalues correspond to
(generalized) states for which the detection time is sharp. The 
eigenstates with complex eigenvalues correspond to states that are 
never detected. 

The time-of-arrival operator is partially characterized by its 
classical limit, which fixes its dependence on the position and 
momenta operators, up to ordering.  We have considered the simple 
case of a nonrelativistic free particle, using a tentative natural 
ordering. A regulation procedure allows us to to find a 
self-adjoint time-of-arrival operator (there is 
no non-detection in this case), and 
we have studied the probabilities the operator yields.  We showed 
that the regulated time-of-arrival operator can be used to derive 
time--energy uncertainty relations, circumventing a well-known 
nonexistence theorem, and that it yields a well-defined ``time 
representation'' for the system.

For more general systems, the classical limit is not likely to be 
sufficient for constructing the operator. In a forthcoming companion 
paper \cite{two} we investigate a general technique for constructing 
the time-of-arrival operator in general cases.  We will study in 
detail the particle in an exponential potential, where the $\hat {\bf 
P}$ operator is non-trivial. We will also investigate parametrized 
systems and theories without a Newtonian time, arguing that the 
ideas presented here may be relevant for the interpretations of 
quantum gravity.

\vskip.3cm

We close with a general comment on the significance of the 
result obtained.  In classical mechanics there is a hidden 
equivalence between the independent time variable $t$ and the 
dependent dynamical variables --the position $x$ in the present 
case.  This equivalence is made manifest by expressing the theory in  
parametrized form,  namely by representing the evolution in terms 
of the functions $t(\tau)$ and $x(\tau)$ of an arbitrary 
parameter $\tau$.  In more elegant and general mathematical 
terms, there are formulations of mechanics, e.g. the presymplectic 
formulation, in which the distinction between dependent and 
independent variables is inessential; see for instance Arnold's 
classic text \cite{arnold}. A parametrized representation is 
commonplace in special relativity (where $T$ is called $x^0$) since 
it allows manifest Lorentz covariance.

It is commonly stated that this equivalence between independent 
(time) and dependent variables is lost is quantum mechanics. The 
arguments in support of this claim are common in the quantum 
gravity literature and take various forms. For instance, it is said 
that the wave function must be normalized by integrating in $x$ and 
cannot be normalized  by integrating in $t$.  Or it is said that 
probabilities are always probabilities of different outcomes 
happening at the same time,  never at the same position. It is our 
impression that these claims are misleading.  The mistake is to 
assume that the $x\leftrightarrow t$ equivalence has to be realized 
as an equivalence in the arguments of the Schr\"odinger wave 
function $\psi(x, t)$. 

The conventional formulation of quantum mechanics in terms of the 
Schr\"odinger wave function $\psi(x, t)$ has {\it already\ } broken 
the $x\leftrightarrow t$ equivalence.  
Indeed, it is a formulation tailored to answer the (experimental) 
question: ``What is the probability of the particle being {\it here 
now}, as opposed to that of being {\it elsewhere now}?''.  The 
corresponding representation diagonalises the Heisenberg operators 
$\hat x(t)$. It is the experimental question considered, and the 
related choice of basis, that breaks the 
$x\leftrightarrow t$ equivalence. Quantum mechanics  allows us to 
consider the following question as well: ``What is the probability of 
the particle getting {\it here now} as opposed to getting {\it here 
elsewhen}?''. In order to answer this question, one is led naturally 
to the $\psi^\pm(T, X)$ representation in which the role of position 
and time are to a large extent interchanged.  In particular, the wave 
function is normalized in time and probabilities of events at the 
same position are considered.

To avoid misunderstandings, let us make clear that we certainly do 
not claim that space and time have the same nature, nor that their 
role in the quantum mechanics of a particle is exactly the same.  
What we suggest is that the common arguments that $x$ and $t$ can 
be treated on equal footing in classical mechanics, but not in 
quantum mechanics, might loose force under closer scrutiny.  
Contrary to the above arguments, our analysis has revealed an 
underlying hidden  equivalence between ``dependent'' and 
``independent'' variables in the quantum theory of a free particle.

\vskip1cm

\noindent
{\bf Acknowledgments} 
\vskip.3cm

\noindent
We thank Ed Gurjoy and Ted Newman for useful
discussions. We are particularly grateful to Don Marolf for the
suggestion of the way to regulate the time-of-arrival operator and 
to Jim Hartle and Jonathan Halliwel for stressing the importance of 
 the problem and for helpful discussions and correspondence.
\newpage

\appendix 
\renewcommand{\theequation}{\Alph{section}.\arabic{equation}}
\section{Is the computed probability density
reasonable ?}

Let us now investigate whether the result we have obtained for the
Newtonian free particle is physically reasonable.  The simplest 
check
is to compute the expectation value of the time-of-arrival of a wave
packet with initial position $x_0$ and momentum $p_0$. The result
should agree with the expected classical time-of-arrival $T=t(X; 
x_0,
p_0)$. Since the operator was constructed via a factor-ordering and
regulation of the classical solution, the expectation values
obviously satisfy Ehrenfest's theorem. A more accurate check which
goes beyond the semiclassical approximation is to compare the
probability distribution we have obtained with the one we can {\it
estimate} by indirect but intuitive methods. In appendix A.1, we 
first
compute an approximation to the probability amplitude 
(\ref{psi(tx)})
--and the resulting probability density-- for
a Gaussian state. This also gives us some intuition for the behavior
of the the distribution. Then in appendix A.2, we compute the
Schr\"odinger current through the detector position, and compare 
this
with the approximation we will obtain in section A.1.

\subsection{Gaussian state}

Consider the Gaussian wavepacket of section 3.5.
In the momentum representation, this state is
given by 
\begin{equation}\label{ss:krep2}
\psi(k)=\left(\frac{2\delta^2}{\pi}\right)^{\frac14} \exp\left(
-(k-k_0)^2\delta^2 -ikx_0\right).
\end{equation}
The envelope of this wave function is a Gaussian of width 
$1/\delta$
at positive momentum $k_0$. In order to simplify the calculations, 
we
slightly modify this state by assuming it to be zero for
$k<\epsilon$. Clearly, since this modification is far out on the tail
of the Gaussian, the error we make is very small (More precisely, 
one
can show that it vanishes as $\exp{(-1/k_0\delta)}$). We thus 
replace
(\ref{ss:krep2}) by
\begin{equation}
\psi(k)=\theta(k-\epsilon)
\left(\frac{2\delta^2}{\pi}\right)^{\frac14} \exp\left(
-(k-k_0)^2\delta^2 -ikx_0\right).
\end{equation}
where $\theta(x)$ is the characteristic function of the positive line. 
Substituting this expression into (\ref{tofxrep}), we find that the 
amplitude
for the particle to be detected between $T$ and $T+dT$ at the 
position $x$ is
\begin{eqnarray}\label{psiapp1}
\psi^+(T, X)&=&\sqrt{\frac\hbar{2\pi
m}}\left(\frac{2\delta^2}\pi\right)^{1/4} 
\int_\epsilon^\infty dk\,\sqrt{k}
\exp\left(-(k-k_0)^2\delta^2 -i\frac{\hbar T}{2 m}k^2
+ik(X-x_0)\right)\cr
&=&\sqrt{\frac\hbar{2\pi
m}}\left(\frac{2\delta^2}\pi\right)^{1/4}\left[
\int_{-\infty}^\infty dk\,\sqrt{k}
\exp\left(-(k-k_0)^2\delta^2 -i\frac{\hbar T}{2 m}k^2
+ik(X-x_0)\right) - R_1\right]\cr&&
\end{eqnarray}
where $R_1$ is again exponentially small, $\sim\exp(-
1/k_0\delta)$, and
we have disregarded an $\epsilon$-dependent phase, which is 
not going to affect the probability distribution.

Now we can expand $\sqrt{k}$ about the peak of the Gaussian as
\begin{equation}\sqrt{k}=\sqrt{k_0}\left(1+\frac{k-k_0}{2k_0}
+ O \left(\frac{k-k_0}{k_0}\right)^2 \right).\end{equation}
Using this expansion, we can compute the Gaussian integral in 
(\ref{psiapp1}) 
\begin{eqnarray}
&&\int_{-\infty}^\infty dk\sqrt{k}
\exp\left(-(k-k_0)^2\delta^2 -i\frac{\hbar T}{2 m}k^2
+ik(X-x_0)\right) \cr
&&=e^{ik_0(X-x_0)-i\frac{\hbar T}{2 m}k_0^2}\left[
\sqrt{\frac{k_0^2\pi}{\delta^2+i\frac{\hbar T}{2 m}}} 
\exp\left[-\frac{(X-x_0-\frac{\hbar Tk_0}{m})^{2}}{4(\delta^2 
+i\frac{\hbar T}{2 m})}\right]\left(1+i
\frac{X-x_0-\frac{\hbar T k_0}{m}}{4k_0(\delta^2 
+i\frac{\hbar T}{2 m})}\right)+R_2\right],\cr&&
\end{eqnarray}
where
\begin{equation} |R_2|\le\frac1{8k_0^{3/2}}\int_{-\infty}^\infty 
dk\, k^2
e^{-k^2\delta^2}=\sqrt{k_0\pi}\frac1{16k_0^2\delta^2}\frac1\delta
\end{equation}
is of second order in $1\over\delta k_0$.  Thus, to first order in the 
(small) quantity  $1\over\delta k_0$, we have 
\begin{equation}
\psi^+(T;x)=\sqrt{\frac{\hbar
k_0}{m}}\left(\frac{\delta^2}{\pi}\right)^{1/4}
e^{ik_0(X-x_0)-i\frac{\hbar T}{2 m}k_0^2}
\frac{\exp\left(-\frac{(X-x_0-\frac{\hbar 
Tk_0}{m})^{2}}{4(\delta^2 
+i\frac{\hbar T}{2 m})}\right)}{4(\delta^2 +i\frac{\hbar T}{2 
m})^{1/2}}
\left(1+i
\frac{X-x_0-\frac{\hbar Tk_0}{m}}{4k_0(\delta^2 
+i\frac{\hbar T}{2 m})}\right),
\end{equation}
To this approximation, the probability density
$\pi^+(T;x)=|\psi^+(T;x)|^2$ is then 
\begin{equation}
\pi^+(T;x)=\frac\hbar{m\sqrt{2\pi}}
\frac{(k_0\delta^2+\frac{(X-x_0)T\hbar}{4m\delta^2}+
\frac{k_0(X-x_0-k_0\hbar T/m)^2}{16k_0^2\delta^2})}
{\left(\delta^2+\frac{T^2\hbar^2}{4m^2\delta^2}
\right)^{3/2}}
\exp\left(-\frac{(X-x_0-k_0T\hbar/m)^2}{2\left(\delta^2+
\frac{T^2\hbar^2}{4m^2\delta^2}\right)}\right).  
\end{equation}
Due to the Gaussian factor, the probability distribution 
is  centered on the classical
time-of-arrival $T=m{X-x_0\over \hbar k_0}$, with width 
$\delta$, and vanishes exponentially outside such a region. 

Notice also that due to the Gaussian factor the third term
in the numerator of the amplitude of the Gaussian is of order 
$(\frac1{\delta k_0})^2$
in all the region where the probability density is not exponentially 
small, so we may rewrite the probability density as
\begin{equation}\label{probappx}
\pi(T;x) =\frac\hbar{m\sqrt{2\pi}}
\frac{(k_0\delta^2+\frac{(X-x_0)T\hbar}{4m\delta^2})}
{\left(\delta^2+\frac{T^2\hbar^2}{4m^2\delta^2}
\right)^{3/2}}
\exp\left(-\frac{(X-x_0-k_0T\hbar/m)^2}{2\left(\delta^2+
\frac{T^2\hbar^2}{4m^2\delta^2}\right)}\right),  
\end{equation}
as an approximation correct to order $(\frac1{\delta k_0})$. 

\subsection{Current}

In ordinary quantum mechanics, one can define a current whose time 
and space components are
\begin{eqnarray}
j^0(x, t) &=& \rho(x, t)=\overline{\psi(x, t)}\psi(x, t) \\ 
j^i(x, t) & = & \frac\hbar{2 m i} (\overline\psi\partial_i\psi-
\psi\partial_i\overline\psi)(x, t),
\end{eqnarray}
where $\psi(x, t)$ is the quantum state in the conventional 
Schr\"odinger-picture position representation.
Since the state satisfies Schr\"odinger's equation, this current is 
conserved
\begin{equation} \partial_0 j^0 + \partial_i j^i =0.  \end{equation}
Consider, for a particle in one dimension, the spacetime region (a
half strip) defined by $x\le X, t_1\le t\le t_2$, and integrate the
divergence free current over this volume. Dropping the boundary 
terms
at $x=-\infty$, we find that the outgoing (i.e., rightward) flux of
this current through the timelike boundary at $x=X$ in the time
interval $(t_1,t_2)$ is
\begin{eqnarray}\label{jxint}
\int_{t_1}^{t_2}dt\> j_{x}&\equiv&\frac\hbar{2 m 
i}\int_{t_1}^{t_2}dt\>
\left. (\overline\psi\partial_x\psi-
\psi\partial_x\overline\psi)\right|_{x=X}\cr
&=&\int_{-\infty}^{x} dx\> \rho|_{t_1}-\int_{-\infty}^{x} dx\> 
\rho|_{t_2}.
\end{eqnarray}
The last line in the above equation is the probability that the
particle is found in the LHS region ($X\le x$) at time $t_1$ minus
the probability that the particle is found in the LHS region at a
later time $t_2$.  It is tempting to identify the flux between the
times $T$ and $T+dT$ (through the timelike surface at $x=X$) of the
current density
\begin{equation}\label{jx} 
j_X(T)dT=\frac\hbar{m}Im\left(\overline{\psi(x,T)}\partial_x\psi(x,T)
\right)_{x=X} dT
\end{equation}
as the probability density that a
detector placed at $x=X$ will detect the particle. Note that the 
Schr\"odinger current density
has the correct dimensions of a density in time, namely 
$[T]^{-1}$. The problem, of course, is that 
the current represents the {\it net\,} flux of
probability across $x=X$, and thus corresponds to the
difference between the probability of crossing to the right and the
probability of crossing to the left.   Namely we may expect that
\begin{equation}
	j_X(T) \sim \pi^+(T; X)-\pi^-(T; X).\end{equation}
within some approximation.   In fact, the 
current $j_X(T)$ is {\it not} positive definite, and we believe that
this is related to the difficulties
 in
the approaches of \cite{hallref} and \cite{nkumar}. 
Equivalently, we may try to interpret the 
current {\it of a pure right moving state\,} as the probability
density that the particle crosses $x=X$.  We are reassured in
doing this by the fact that in this case the current is 
positive definite, and its integral over all times gives one. 
This conclusion is based on the assumption that
the particle cannot ``zigzag'' across the $X=x$ line, an
assumption which might be valid only if we look at sufficiently
large times. 

So, when $\psi$ is a pure right moving state, the rightward flux 
density
is positive%
\footnote{See (\ref{jx}) and recall that this is a free particle.} 
and integrates to 1 over all time. (To see this, take the limits
$t_1\rightarrow -\infty$ and $t_2\rightarrow \infty$ in
(\ref{jxint}).)
It is therefore at
least consistent to interpret this as an estimate of the  probability
density. Does this
estimate yield the correct semi-classical limit? The
expectation values of the usual
position and momentum operators satisfy Ehrenfest's theorem, since 
the
probability densities are associated with the decomposition of a 
state
onto the spectrum of some self-adjoint operator. The same is true of
the probability densities we have obtained from the time-of-arrival
operator. The above
current density is {\em not} obtained via a spectral projection, 
however,
and is {\em not} associated with a ``time operator''. How do the
expectation values of the time-of-arrival behave wrt. this 
estimated probability density, and do they
correspond to the classical limit in some way? We next proceed to
analyze this issue. 

The expectation value of the time-of-arrival of the particle at the
position $X$ is naturally defined only when the state has support 
only
on the $k\ge0$ region and is then given by
\begin{eqnarray}
\ev{T}_X&=&\int_{-\infty}^\infty dT\> T\, j_X(T)\cr
&=&\frac\hbar{2\pi m}Re\left\{\int_{-\infty}^\infty 
dT\int_0^\infty dk
\int_0^\infty dk'\> T\cdot \exp\left(\frac{i\hbar T
k^2}{2m}-ikX\right)\overline{\psi(k)} \exp\left(\frac{-i\hbar T
k'^2}{2m}+ik'X\right)\psi(k') \right\}\cr
&=&\frac\hbar{2\pi m}Re\left\{\int_{-\infty}^\infty 
dT\int_0^\infty dk
\int_0^\infty dk'\>
\frac{m}\hbar\frac1{k}\left[\left(\frac1{i}\frac\partial{\partial k}+X\right)
\cdot \exp\left(\frac{i\hbar T
k^2}{2m}-ikX\right)\right]\right.\times\cr
&&\left.\hphantom{aaaaaaa}\overline{\psi(k)} \exp\left(\frac{-
i\hbar T
k'^2}{2m}+ik'X\right)\psi(k') \right\}\cr
 &=& \frac{m}\hbar Re\left\{\int^\infty_0 dk\> 
\overline{\psi(k)}\left( 
\frac{X}{k} - \frac{i}{k} \frac\partial{\partial k}\right)\psi(k)\right\}\cr
&\equiv& mX\left\langle\frac1{\hat p_0}\right\rangle - \frac{m}2
\left\langle\frac1{\hat p_0}\hat x_0 
+ \hat x_0\frac1{\hat p_0}\right\rangle,
\end{eqnarray}
where we have dropped a surface term at $k=0$ since we assume 
that
$\psi(k)_{k\rightarrow0}\rightarrow k^{2+}$. The final expression in
the above equation is the symmetric factor ordering of the 
classical expression $T=m(X-x_0)/p_0$, thus we do recover the
correct semiclassical limit. 

For the localized right moving wave packet 
previously considered, the current is easily computed, giving
\begin{equation}
j_X(T)=\frac\hbar{2 \pi m} Re\left[\int^\infty_0
dk \exp\left( \frac{iT\hbar k^2}{2 m}-ikX
\right)\overline{\psi(k)} \cdot  \int^\infty_0
dk' k'\,\exp\left( \frac{-iT\hbar k'^2}{2
m}+ikX\right)\psi(k')\right].
\end{equation}
The integral can be explicitly done, yielding 
\begin{equation}
j_X(T)=\frac\hbar{m\sqrt{2\pi}}
\frac{(k_0\delta^2+\frac{(X-
x_0)T\hbar}{4m\delta^2})}{\left(\delta^2+\frac{T^2\hbar^2}{4m^2\delta^2}
\right)^{3/2}}
\exp\left(-\frac{(X-x_0-k_0T\hbar/m)^2}{2\left(\delta^2+
\frac{T^2\hbar^2}{4m^2\delta^2}
\right)}\right).  
\end{equation}
which is precisely the approximate form for the probability we
computed from the time-of-arrival operator (see
(\ref{probappx})). Thus, the probability distributions computed with
the time-of-arrival operator and by means of the current agree for a
right moving localized wave packet to order $(1/ k_0\delta)$, which 
is one order beyond the classical limit.

For a general state, roughly 
localized in momentum state around a momentum $k_0$, 
we can compare the
current (\ref{jx}) with the difference between the probability
of being detected moving towards the right minus the probability
of being detected moving left.  Namely we can estimate
\begin{equation} d_X(T) = j_X(T) - ( \pi_X^+(T)-\pi_X^-(T)). 
\end{equation}
Explicitly, using (\ref{jx}) and (\ref{probdensT3}), we have
\begin{eqnarray}
d_X(T) &=& 
\frac\hbar{m}Im\left(\overline{\psi(X;T)}\partial_X\psi(X;T)\right)_{x=X}
 - \frac\hbar{2 \pi m} \left|\int_{0}^{+\infty}
  dk \sqrt{k}\exp\left(\frac{-iT\hbar k^2}{2
m}+ikX\right)\psi(k) \right|^2\cr 
&&+  \frac\hbar{2 \pi m} \left|\int_{-\infty}^{0}
  dk \sqrt{k}\exp\left(\frac{-iT\hbar k^2}{2
m}+ikX\right)\psi(k) \right|^2 .
\end{eqnarray}
We now put $X=0$ for simplicity, without loss of generality. 
A little algebra gives 
\begin{equation} d_0(T) = \frac\hbar{2 \pi m} \int_{-
\infty}^{+\infty}
  dk \int_{-\infty}^{+\infty} dk' (\sqrt{k}-\sqrt{k'})^2 
\exp\left(\frac{-iT\hbar (k^2-k'{}^2)} 
{2m}\right)\psi(k)\overline{\psi(k')} .
\end{equation}
Now suppose that our measurement has a precision $\delta T$. Then 
the difference of the probability densities, averaged around $T$ is
\begin{equation}
\Delta P(T, \delta T)
 = \frac1{2 \delta T}\int_{T-\delta T}^{T+\delta T}dT'\>d_0(T').
\end{equation}
The integral in $dT'$ can be done easily, which yields
\begin{equation}
\Delta P(T, \delta T) = \frac\hbar{2 \pi m}  \int_{-\infty}^{+\infty}
  dk \int_{-\infty}^{+\infty}
  dk' (\sqrt{k}-\sqrt{k'})^2 e^{\left(\frac{-iT\hbar (k^2-k'{}^2)}
{2m}\right)}\psi(k)\overline{\psi(k')} \left\{\frac{\sin(
\frac{\delta T\hbar}{2m}(k^2-k'{}^2)) }{
\frac{\delta T\hbar}{2m}(k^2-k'{}^2) }\right\}.
\end{equation}
For large $\delta T$, compared to the ``deBroglie time'' 
of the particle $\frac{2m}{k^2\hbar}$ (where $k$ is the highest 
momentum
in the support of the wave function), the factor $\{\,\}$ in the
integrand approaches a delta function  
in the difference between the
two integration variables (plus a term with a delta function in the
sum of the two integration variables, which we may assume to be
negligible for states sufficiently localized in momentum space), 
and the integral is then suppressed 
by the $(\sqrt{k}-\sqrt{k'})^2 $ factor. This indicates that the
two ways of computing the probability for the time-of-arrival
approach each other when our resolution time is larger than the
particle's deBroglie time.

\end{document}